\documentclass{aa}

\usepackage{lmodern}
\usepackage{amsmath}
\usepackage{cancel}
\usepackage{todo}
\usepackage{wasysym}
\usepackage{upgreek}
\usepackage{txfonts}
\usepackage[T1]{fontenc}
\usepackage[utf8]{inputenc}
\usepackage{lmodern}
\usepackage{ulem}
\usepackage{xcolor}

\graphicspath{ {files/} }

\begin{document}

\title{Angular differential kernel phases}
\author{Romain Laugier \inst{1}
        \and
        Frantz Martinache \inst{1}
        \and
        Nick Cvetojevic \inst{1}
        \and
        David Mary \inst{1}
        \and
        Alban Ceau \inst{1}
        \and
        Mamadou N'Diaye \inst{1}
        \and
        Jens Kammerer \inst{2,7}
        \and
        Julien Lozi \inst{3}
        \and
        Olivier Guyon \inst{3,4,5,6}
        \and
        Coline Lopez \inst{1}
        }
\institute{
        Université Côte d'Azur, Observatoire de la Côte d'Azur, CNRS, Laboratoire Lagrange, France
        \and
        Research School of Astronomy \& Astrophysics, Australian National University, ACT 2611, Australia
        \and
        Subaru Telescope, National Astronomical Observatory of Japan, National Institutes of Natural Sciences (NINS), 650 North Aohoku Place, Hilo, HI, 96720, U.S.A.
        \and
        Steward Observatory, University of Arizona, Tucson, AZ, 85721, U.S.A.
        \and
        College of Optical Sciences, University of Arizona, Tucson, AZ 85721, U.S.A.
        \and
        Astrobiology Center of NINS, 2-21-1, Osawa, Mitaka, Tokyo, 181-8588, Japan
        \and
        European Southern Observatory, Karl-Schwarzschild-Str 2, 85748, Garching, Germany
        }

\abstract
{To reach its optimal performance, Fizeau interferometry requires that we work to resolve instrumental biases through calibration. One common technique used in high contrast imaging is angular differential imaging, which calibrates the point spread function and flux leakage using a rotation in the focal plane.
}
{Our aim is to experimentally demonstrate and validate the efficacy of an angular differential kernel-phase approach, a new method for self-calibrating interferometric observables that operates similarly to angular differential imaging, while retaining their statistical properties.
}
{We used linear algebra to construct new observables that evolve outside of the subspace spanned by static biases. On-sky observations of a binary star with the SCExAO instrument at the Subaru telescope were used to demonstrate the practicality of this technique. We used a classical approach on the same data to compare the effectiveness of this method.
}
{The proposed method shows smaller and more Gaussian residuals compared to classical calibration methods, while retaining compatibility with the statistical tools available. We also provide a measurement of the stability of the SCExAO instrument that is relevant to the application of the technique.
}
{Angular differential kernel phases provide a reliable method for calibrating biased observables. Although the sensitivity at small separations is reduced for small field rotations, the calibration is effectively improved and the number of subjective choices is reduced.
}

\maketitle

\section{Introduction}
  Since the advent of speckle interferometry \citep{Labeyrie1970}, Fizeau interferometry techniques that use the aperture of a single telescope have proven to be a reliable way to obtain measurements at and beyond the classically defined resolution limit of telescopes. By working with the Fourier transform of images, they exploit observables that were originally developed for long baseline interferometry, such as closure phases \citep{Jennison1958, Baldwin1986}, to provide observables that are robust for instrumental phase errors. Non-redundant masking, in particular, has been established as an observing mode in most of the high-resolution instruments available \citep{Tuthill2010}. Kernel-phase observables \citep{Martinache2010} rely on a generalization of the notion of closure phase for redundant apertures that is applicable in the high Strehl regime.\par
  These interferometric techniques rely on simplifications, such as the monochromatic approximation, the short exposure approximation, or the absence of scintillation and instrumental amplitude errors. As described by \citet{Ireland2013}, deviations from these model conditions induce biases in the data that may sometimes prevent direct interpretations. Calibration observations, where calibrator targets are observed in the same conditions as the science target, are routinely used to remove these instrumental biases.\par
  Techniques exist for improving the performance of the calibration, such as optimal calibration weighting \citep{Ireland2013}, or Karhunen-Loève projection \citep{Kammerer2019}. In both of those approaches, a number of different calibrators are used either to better interpolate the calibration signal or to project the target signal outside of the subspace containing the most common calibration signals. Both of these methods attempt to circumvent the problem of hidden variables that affect the quality of the calibration, but they tend to be limited by the small number of calibration sources available and the diversity they represent. Sampling this diversity requires the acquisition of a lot of data on calibration sources which is very inefficient. Among the hidden variables are spectral properties of the targets, inaccuracies of the model, and quasi-static instrumental biases. An ideal solution to the problem would be, therefore, to have a method to calibrate the science target with itself during the same observing session, thus preserving the consistency of the aforementioned parameters.\par
  Systematic biases also affect coronagraphic observations, and similar approaches are used to characterize the flux leakage through coronagraphs. Angular differential imaging (ADI), proposed by \citet{Marois2006a}, uses the opportunity provided by the fact that the telescope uses an alt-azimuth mount and the relay train provides a fixed pupil relative to the detector along with a slowly rotating field of view. In consequence, the technique is capable of distinguishing instrumental errors that are fixed with the pupil from features of the target that are rotating with the field. The technique has been used extensively as part of different analysis processes such as TLOCI \citep{Marois2014}, KLIP \citep{Soummer2012}, or PYNPOINT \citep{Amara2012}.\par
  The application of such a technique to closure phases or kernel phases has, to our knowledge, not yet been reported. We therefore introduce a mathematical expression for angular differential kernels (ADK), a self-calibration technique that modifies the observables to completely remove the static bias while preserving their properties, both in terms of instrumental phase rejection and in terms of versatility, therefore building upon the current tools that have been developed for kernel-phase analyses.
  
\section{Analytical formulation}
In this section, we describe a mathematical framework for generating self-calibrated ADKs building up from the traditional kernel-phase formalism. 

  Subsection \S \ref{sect:classical} provides the context for kernel-phase observables and discusses the nature of the biases that arise. Construction of the new self-calibrating observables is handled in two steps: construction of the projection and whitening.

\subsection{Classical calibration}\label{sect:classical}

  The fundamental approximation of kernel-phase techniques is based on the finding that at high Strehl ratios (behind ExAO or in space), the phase error in the telescope's Pupil-plane $\boldsymbol{\varphi}$ will, to first-order, propagate linearly into the Fourier plane phase \citep{Martinache2010}. This linear application is described by the matrix $\mathbf{A}$, which leads, through the van Cittert-Zernike theorem, to a comparison of the observations and the model as follows:
    \begin{equation}\label{eq:basic_phase_fit}
      \boldsymbol{\Phi} = \boldsymbol{\Phi}_0 + \mathbf{A} \cdot \boldsymbol{\varphi} + \boldsymbol{\varepsilon}'\,
    ,\end{equation}
where $\boldsymbol{\Phi}$ and $\boldsymbol{\Phi}_0$ are the vectors of observed and theoretical Fourier phases and $\boldsymbol{\varepsilon}'$ are deviations between model and observation. Their dimension is the number of considered baselines defined by a discrete representation of the pupil. Identifying the left null space of $\mathbf{A}$ \citep{Martinache2010} provides the matrix $\mathbf{K}$:

  \begin{equation}
    \mathbf{K} \cdot \mathbf{A} \cdot \boldsymbol{\varphi} = 0
  ,\end{equation}
  therefore leading to the model-fitting equation:
  \begin{equation}\label{equ:model-fitting}
    \mathbf{K} \cdot \boldsymbol{\Phi} = \mathbf{K} \cdot \boldsymbol{\Phi}_0 +  \mathbf{K} \cdot \boldsymbol{\varepsilon}'
  .\end{equation}
The origins of different error sources that comprise $\mathbf{K}\cdot \varepsilon'$ have previously been described by \citet{Ireland2013} and are also described further below. To simplify our notations, we introduce $\boldsymbol{\kappa} = \mathbf{K} \cdot \boldsymbol{\Phi} $ as the observables, $\boldsymbol{\kappa}_{0} = \mathbf{K} \cdot \boldsymbol{\Phi}_0$ as the model. For the purpose of this work, we decompose the errors into two types based on their temporal properties, namely, $\mathbf{K} \cdot \boldsymbol{\varepsilon}' =  \boldsymbol{\kappa}_{bias}  + \boldsymbol{\varepsilon}$, where $\boldsymbol{\kappa}_{bias}$ regroups the errors that remain stable over the timescale of the observations (the bias) and $\boldsymbol{\varepsilon}$ regroups all the errors varying during the observation. Therefore, the model-fitting equation is written as follows:
  \begin{equation}\label{eq:simple_kernel}
    \boldsymbol{\kappa} = \boldsymbol{\kappa}_{0} + \boldsymbol{\kappa}_{bias}  + \boldsymbol{\varepsilon}.
  \end{equation}
The varying error can encompass both correlated effects of sensor noise (readout noise and shot noise) or by third-order residual effects of wavefront error leaking through the kernel matrix. Here, both are treated as Gaussian \citep{Ceau2019b} and described by a covariance matrix $\boldsymbol{\Sigma}$. \par
This covariance is usually estimated through either analytic propagation of the shot noise \citep{Kammerer2019}, image-plane bootstrapping of the shot noise \citep{Ceau2019b, Laugier2019}, or empirically, when a sufficient number of short exposure images is available as in the case of the example presented here. The covariance $\boldsymbol{\Sigma}$ is used to construct a square whitening matrix 
\begin{equation}\label{equ:whitening_matrix}
    \mathbf{W} = \boldsymbol{\Sigma} ^{-\frac{1}{2}},
\end{equation}
corresponding to a subspace where the considered errors are uncorrelated. This transforms Equation (\ref{eq:simple_kernel}) into:
  \begin{equation}
    \mathbf{W}\boldsymbol{\kappa} =
    \mathbf{W} (\boldsymbol{\kappa}_{0} + \boldsymbol{\kappa}_{bias} + \boldsymbol{\varepsilon}) ,
  \end{equation}{}
  which decorrelates uncertainties ($\mathrm{Cov}(\mathbf{W}\mathbf{\varepsilon}) = \mathbf{I}$),  where $\mathbf{W}$ is the whitening matrix.\par
  
  This approach, which lies at the heart of our statistical treatment, allows for optimal performance if the evaluation of the covariance is representative of the actual experimental errors.\par

  A calibration signal $\boldsymbol{\kappa}_{calibrator}$  acquired on an unresolved source ($\boldsymbol{\kappa}_{0} = 0$) is subtracted from the target signal as follows:
  \begin{equation}
   \mathbf{W}(\boldsymbol{\kappa}_{target} - \boldsymbol{\kappa}_{calibrator}) = \mathbf{W}(\boldsymbol{\kappa}_{0} + \boldsymbol{\varepsilon}'') ,
  \end{equation}
  In this manner, the biases $\boldsymbol{\kappa}_{bias}$, which are present in both observations, are canceled, whereas $\boldsymbol{\varepsilon}''$ now contains the combined errors from the target and the calibrator, and $\mathbf{W}$ must therefore be calculated based on the covariance $\boldsymbol{\Sigma}$ that is written as:
  \begin{equation}
      \boldsymbol{\Sigma} = \boldsymbol{\Sigma}_{target} +  \boldsymbol{\Sigma}_{calibrator}
  ,\end{equation}
with $\boldsymbol{\Sigma}_{target}$ and $\boldsymbol{\Sigma}_{calibrator}$ as the covariances of the target and the calibrator, respectively.\par

\subsection{Discussion of the classical approach}
  It should be noted here that $\boldsymbol{\varepsilon}''$ also contains residual calibration errors as described by \citet{Ireland2013} that arise from a number of factors including some intrinsic to the choice of calibrators. Furthermore, we make the assumption that the calibrator is an unresolved point source, so that $\boldsymbol{\kappa}_{0}$ only represents the signal from the target. In practice, verifying this hypothesis is difficult, especially when the observing campaign has cutting-edge sensitivity and attempts to make new discoveries. Thus, any resolved signature in the calibrator propagates through the entire pipeline and either imprints artifacts or false detections on the final results, or it suppresses a real companion signal.
  While a valid approach is to use a second calibrator to lift the ambiguity, requiring two calibrators for each target is inefficient as a smaller fraction of observation time is spent on the target.\par
  Experience shows that observations of different calibration sources differ by amounts larger than can be explained by the statistics observed for each \citep{Kraus2008, Martinache2009, Laugier2019}; this is either because of the differences in spectral properties between the calibrators or because of slow variations in observing conditions. For this reason, a larger number of calibration sources are often used to try to improve the performance of the calibration step as described by \citet{Ireland2013} and \citet{Kammerer2019}.\par
  Unfortunately, the distribution of the calibration errors are very difficult to evaluate both because of the small number of realizations (i.e., of calibrators) and because they combine variations of the spectral properties of the target, sky location leading to different air mass, adaptive optics performance, and telescope flexures. Our approach mitigates this effect while preserving the possibility to use the whitening transform to manage the fast varying errors.

\subsection{Angular differential kernels}\label{sect:generalization}
  An alternative approach would be to completely remove the errors introduced by the intrinsic differences between target and calibrator by using only observations of the target itself in the logic of self-calibration. As is the case in ADI, the signal of interest would then be identified through the diversity brought on by the field rotation.\par
  For this purpose, we consider an arbitrary $n_f$ number of frames acquired during an observing sequence, each taken at a known field position angle (parallactic angle for ground based telescopes, or roll angle for space telescopes, or other).\par
  We consider a series of $n_f$ kernel-phase observables vectors $\boldsymbol{\kappa}_i$ extracted from the series of images, each corresponding to a different field rotation angle. We concatenate the whole dataset into a single long observable vector $\boldsymbol{\kappa}_s$ of length $n_k \times n_f$, where $n_k$ is the number of kernel-phase observables. The model observables $\boldsymbol{\kappa}_{0,s}$ and the residuals $\boldsymbol{\varepsilon}_s$ are treated in the same way, leading to the new model-fitting equation, now written with the concatenated observables:
  \begin{equation}
    \boldsymbol{\kappa}_s = \boldsymbol{\kappa}_{0,s} + \boldsymbol{\kappa}_{bias,s} + \boldsymbol{\varepsilon}_s,
  \end{equation} 
  where $\boldsymbol{\kappa}_{bias,s}$ is the contribution of the static bias on all the observable vectors, and $\boldsymbol{\varepsilon}_s$ is the residual in this concatenated form. Since it is the same for all the frames, this contribution can be expressed as a function of an arbitrary single  kernel-phase signal of dimension $n_k$:
  \begin{equation}
      \boldsymbol{\kappa}_{bias,s} =\mathbf{U}_f \cdot \boldsymbol{\kappa}_{bias} ,
  \end{equation}
  where $ \boldsymbol{\kappa}_{bias} $ is the static calibrator signal of length $n_k$, and $\mathbf{U}_f$ is an unfolding matrix that maps the constant calibrator signal into a series of repeated signals (a concatenation of $n_f$ identity matrices $\mathbf{I}$, each of size $n_k$ by $n_k$)\par
  \begin{equation}
    \mathbf{U}_f = 
    \begin{bmatrix}
        1 & 0 & \cdots & 0 \\
        0 & 1 & \cdots & 0 \\
        \vdots & \vdots & \ddots & \vdots \\
        0 & 0 & \cdots & 1 \\
        \vdots & \vdots & \vdots & \vdots\\
        \vdots & \vdots & \vdots & \vdots\\
        1 & 0 & \cdots & 0 \\
        0 & 1 & \cdots & 0 \\
        \vdots & \vdots & \ddots & \vdots \\
        0 & 0 & \cdots & 1 \\
    \end{bmatrix}
    =
    \begin{bmatrix}
        \mathbf{I} \\
        \mathbf{I} \\
        \vdots \\
        \mathbf{I} \\
    \end{bmatrix}.
  \end{equation}
  The model-fitting equation can then be formulated as follows:
  \begin{equation}\label{equ:uf}
    \boldsymbol{\kappa}_s = \boldsymbol{\kappa}_{0,s} + \mathbf{U}_f \cdot \boldsymbol{\kappa}_{bias} + \boldsymbol{\varepsilon}_s,
  \end{equation}
  where any evolution of the quasi-static component has to be considered as part of $\boldsymbol{\varepsilon}_s$. In a manner reminiscent of the ADI approach, we build a matrix $\mathbf{L}$ that cancels out the static bias in the observables. This matrix subtracts from each of the observable vectors $\boldsymbol{\kappa}_i$, the mean of the observable vectors of each frame. We call this matrix $\mathbf{L}$ and express it as:
  \begin{equation}\label{Lmatrix}
    \mathbf{L} = \Big( \mathbf{I} - \frac{1}{n_f}\mathbf{U}_f  \mathbf{U}_f^\mathrm{T} \Big)\,.
  \end{equation}
  Details behind the construction of this matrix and its properties can be found in appendix \ref{app:projection}. Since, by design, $\mathbf{L}\mathbf{U}_f = 0$, multiplying by $\mathbf{L}$ brings to zero any purely static contribution to the signal. Although the signal of interest $\boldsymbol{\kappa}_{0,s}$ may contain some static part that will be lost, the antisymmetric nature of phase signatures relative to the image plane guarantees that any feature with a non-zero phase signature will never be perfectly static with field rotation. As a consequence, some sensitivity loss is expected at small separations and small field rotations, the same as ADI.\par

\subsection{Statistical whitening}\label{sect:whitening}
  The fact that the application described by $\mathbf{L}$ is not a surjection poses problems to subsequent statistical treatment as it implies a singular covariance for the resulting observables. In order to make direct use of statistical tools, we use a modified version $\mathbf{L}'$ of the matrix $\mathbf{L}$ , which spans the same subspace but has been made into a surjection by removing its last $n_k$ lines.\par
  This turns Equation (\ref{equ:uf}) into:
  \begin{equation}\label{equ:lp_uf}
    \mathbf{L}' \boldsymbol{\kappa}_s = \mathbf{L}' \boldsymbol{\kappa}_{0,s}  + \mathbf{L}' \boldsymbol{\varepsilon}_s.
  \end{equation}
  This is better described in appendix \ref{app:dim_reduction}; also, it does not affect the information gathered by the observables.\par 
  In the particular case $ n_f = 2$, the $\mathbf{L}'$ matrix can be written as two blocks:
  \begin{equation}\label{equ:pairwise_matrix}
      \mathbf{L}' = 
      \begin{bmatrix}
        \frac{1}{2}\mathbf{I} & - \frac{1}{2}\mathbf{I}
      \end{bmatrix}
      ,
  \end{equation}
  which reveals that it operates a subtraction of the observables in the case of two frames. This algebraic approach provides flexibility of the number of frames that are acquired during a session, as well as the moment they are taken.\par
  Assuming the statistical errors affecting the measurements are independent from one frame to the other, the covariance matrix of the concatenated observables is block-diagonal:
  \begin{equation}
    \boldsymbol{\Sigma}_s = 
    \begin{bmatrix}
        \boldsymbol{\Sigma}_1 & 0 & \cdots & 0\\
        0 & \boldsymbol{\Sigma}_2 & \cdots & 0\\
        \vdots & \vdots & \ddots & \vdots\\
        0 & 0 & \cdots &  \boldsymbol{\Sigma}_{n_f}\\
    \end{bmatrix}
  ,\end{equation}
  where $\boldsymbol{\Sigma}_i$ is the covariance matrix of the $i^{th}$ kernel-phase observable vector. Therefore, the covariance of the ADK observables built with the matrix $\mathbf{L'}$ becomes:
  \begin{equation}
    \boldsymbol{\Sigma}_{ADK} = \mathbf{L'} \boldsymbol{\Sigma}_s \mathbf{L'}^T
  .\end{equation}
  In this case, the covariance matrix is invertible, and a whitening matrix $\mathbf{W}_{ADK}$ can be computed with Equation (\ref{equ:whitening_matrix}), leading to the model-fitting equation to be used in the data reduction:
  \begin{equation}\label{equ:model-fitting_reduction}
    \mathbf{W}_{L}\mathbf{L}' \boldsymbol{\kappa}_s =
    \mathbf{W}_{L}\mathbf{L}' \boldsymbol{\kappa}_{0,s}  + \mathbf{W}_{L}\mathbf{L}' \boldsymbol{\varepsilon}_s.
  \end{equation}
  In line with the goal of this work, this equation can be used in all the typical applications of kernel phases, such as model-fitting and hypothesis testing. Observable quantities that are thus formed are insensitive to static biases and make up a generalization of simpler intuitive approaches. While similar in principle to what is carried out with ADI in the image plane, their covariance can be inverted for more rigorous post-processing techniques.\par
  An alternative approach consists in deriving the same reasoning on a whitened version of Equation (\ref{equ:uf}). We show in appendix \ref{app:post_whitening_proj} that this approach, although it is more complicated mathematically, leads to a projection matrix that is equivalent.

\section{On-sky validation}
  Having established the mathematical principles for a self-calibration procedure, we now propose a practical implementation and on-sky demonstration of its usability to show that an observing sequence of a rotating field can be exploited without the need to observe a calibration source. A classical calibration approach will be followed in parallel to provide a qualitative comparison.\par
  Although the approach can also be used with other kinds of observables, we use full-pupil kernel-phase observations, which requires an imaging instrument that provides a good wavefront correction.

\subsection{The SCExAO instrument}
  The Subaru Coronagraphic Extreme Adaptive Optics (SCExAO) \citep{Jovanovic2015a,Lozi2018} instrument is a planet-hunting, high-contrast imaging instrument on the Subaru telescope located on Maunakea, Hawaii. It is equipped with a 50x50 actuators deformable mirror controlled in closed-loop with a pyramid wavefront sensor at several kHz. It operates as a second stage to the facility adaptive optics system AO188 which corrects 188 modes with a larger stroke using a curvature wavefront sensor.\par
  Our observations took place within some of the dedicated on-sky engineering time of the instrument, in parallel with other characterization and commissioning tasks, which was decisive in the choice of targets and observing time. We used the internal near-infrared camera without any coronagraphic mask. This camera is a C-RED2 camera providing low readout noise (30 electrons) at high frame rates (up to several kHz in cropped mode). It provides a plate scale of 16.2 mas per pixel, which satisfies the Nyquist-Shannon sampling requirement in the H band.\par
  The non-common-path aberrations (NCPA) on SCExAO are generally corrected for the CHARIS module \citep{Groff2015}, which is the prime near-infrared scientific instrument for high-contrast observations. As a consequence, the data was affected by large amounts of low-order static phase error dominated by astigmatism and trefoil for a total of $~2$ radians peak-to-valley in the Fourier phase, which is very challenging for kernel-phase observables.\par
\begin{figure}
      \centering
      \includegraphics[width=0.5\textwidth]{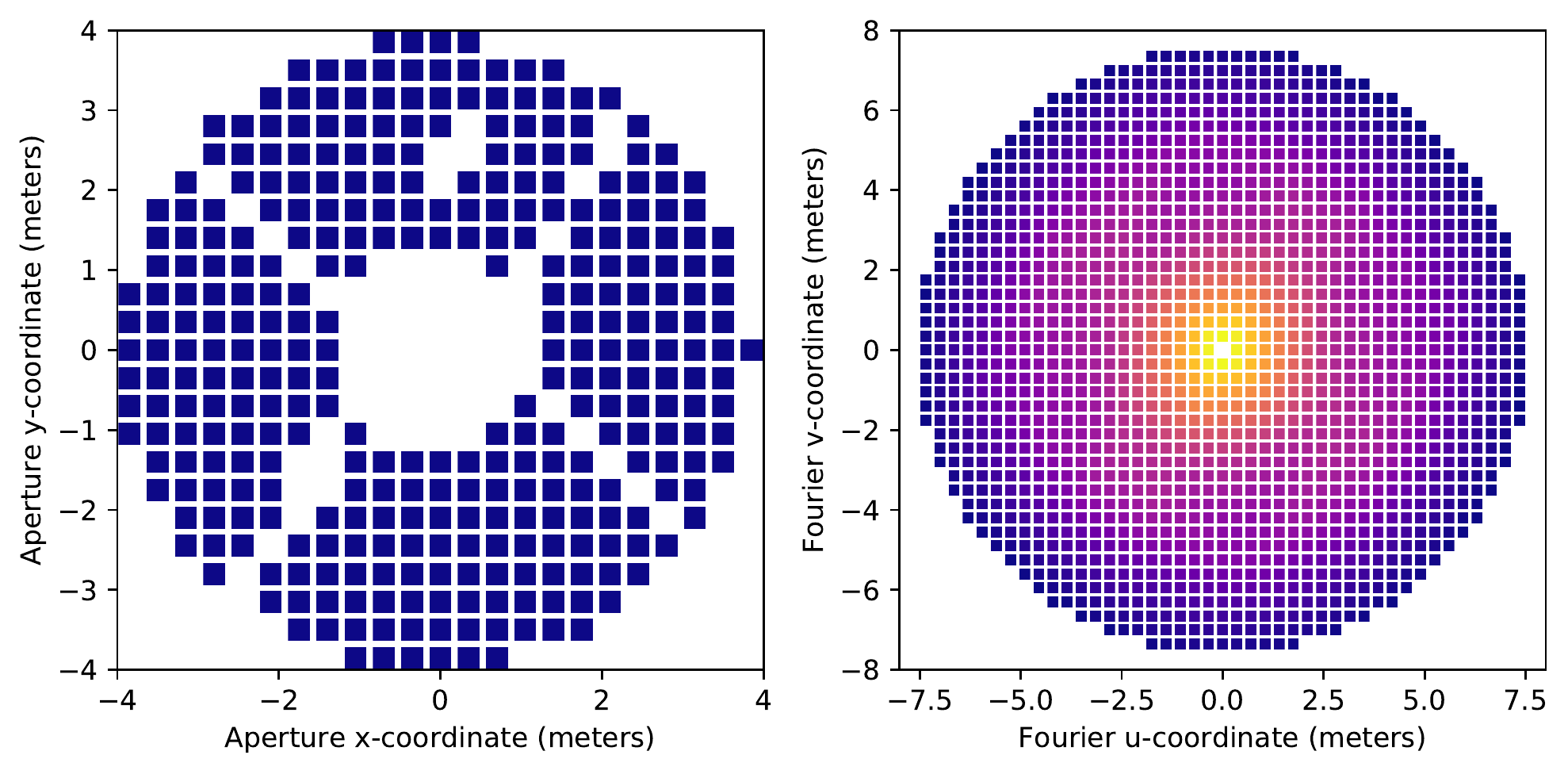}
      \caption{Discrete representation of the pupil of the SCExAO instrument (left) and corresponding uv plane samples (right) used for the analysis of 3 Ser and the calibrator 31 Boo. Some more of the longest baselines were discarded for HD 211976, leading to a slightly smaller sampled uv plane.}
      \label{fig:discrete_model}
  \end{figure}

  The camera was also used to capture the pupil of the instrument using the internal calibration source. The pupil is then discretized using the tools provided in the \verb+xara+\footnote{https://github.com/fmartinache/xara} package with a step of 35cm. This model, represented in Fig. \ref{fig:discrete_model}, is composed of 329 sub-apertures of equal transmission that normally generate 781 baselines to be sampled in the uv plane. However, in order to avoid the problem of $\pi$ radians degeneracy of the phase that appears as a combination of the large NCPA and the phase noise, the longest baselines were discarded. The values of these parameters, and others described in this section, can be found in Table \ref{tab:extraction_parameters}.
  
  {}

  \begin{table}[]
    \centering
    \begin{tabular}{c|c|c|c}
        Parameter            & 3 Ser & 31 Boo & HD 211976 \\
        \hline
        Exposure time (ms)      &0.308 &0.308 & 1.99 \\
        Image domain coadd & 2    & 2    & 4    \\
        Selection ratio (\%)    & 5    & 5    & 50   \\
        Field rotation bin (deg)& 1.8  &-     & 1.8  \\
        Number of valid bins    & 16   &1     & 23   \\
        Total field rotation (deg)& 29.& 0.   & 40.  \\
        \hline
        Maximum baseline (m) & 7.6 & 7.6 & 7.3\\
        Number of  baselines    &  740  & 740 & 680  \\
        Number of kernel phases &  412  & 412 & 352  \\
        
    \end{tabular}
    \caption{Value of relevant parameters for the extraction of the kernel phases}
    \label{tab:extraction_parameters}
  \end{table}{}

\subsection{The data reduction pipeline}\label{sect:reduction}
  The fast frame rate of the camera resulted in millions of images. All were dark-subtracted and flat-fielded using dome flats. Images were co-added to increase the signal-to-noise ratio (S/N), then the frame selection was applied based on a criterion of empirical Strehl ratio, while the image timestamps were tracked. The number of co-adds and frame selection rates  were selected to prevent the $\pi$ radians degeneracy mentioned before, yet to keep a large enough number of images for the statistical treatment to be valid.\par
  Each frame was then re-centered on the brightest speckle to integer pixel. A fourth-order exponential mask was numerically applied to the images both to reduce the contribution of the read noise in the unused parts of the image and to avoid the Fourier space aliasing \citep{Laugier2019}. Its radius $r_0$ was chosen to correspond to a radius of $0.5 \lambda/b$, where b is the smallest baseline.\par

  The complex visibilities were extracted for the pupil model using the \verb+xara+ package, which computes a discrete Fourier transform at the exact uv coordinates of the model, directly providing a vector of complex visibilities. A gradient descent algorithm was used to find a phase wedge that minimizes the square of the phase in the uv plane, thus providing a subpixel centering. The phase was then multiplied by the kernel matrix to obtain kernel-phase observables.\par
  The parallactic angle was determined for each frame and the kernel phases were binned together with a resolution of a few degrees of field rotation into $n_f$ individual chunks. The value must be selected to be smaller than the rotation angle at which the kernel-phase signal decorrelates, which depends on the extent of the region of interest. Larger separations therefore require smaller angular bins. This allows us to consider the signal as stationary for the statistical analysis, as well as to keep the calculation within reasonable scales. For the bins containing a number of realizations at least three times larger than $n_k$, the mean  $\boldsymbol{\kappa}_i$ and its covariance $\boldsymbol{\Sigma}_i$ are empirically determined. The other bins are discarded so that they do not produce a biased covariance matrix. The mean parallactic angle is also evaluated for each bin to be used in the model fitting.\par
  As described in \S \ref{sect:generalization} yielding $\boldsymbol{\kappa}_s$ and $\mathbf{W}_s$ to be used in what will be subsequently referred to as the raw analysis:
  \begin{equation}\label{equ:raw}
    \mathbf{W}_s \boldsymbol{\kappa}_s = \mathbf{W}_s \boldsymbol{\kappa}_{0,s} + \mathbf{W}_s\boldsymbol{\varepsilon}_{s},
  \end{equation}
  which neglects any bias, where $\mathrm{Cov}(\mathbf{W}_s\boldsymbol{\varepsilon}_{s}) = \mathbf{I}$, and where $\boldsymbol{\kappa}_{0,s}$ is a concatenated signal model evaluated for each of the mean field rotation parallactic angles.\par
  The projection matrix is built based on Equations (\ref{eq:l2.1}), (\ref{eq:l2.2}), and (\ref{eq:l2.crop}) and to be used in the ADK model-fitting Equation (\ref{equ:model-fitting}).\par
  For the standard calibration strategy, the mean observables of the calibrator $\boldsymbol{\kappa}_{calibrator}$ were subtracted from each of the target observables and the covariance adjusted accordingly as the sum of the covariance matrices of the two signals, leading to the corresponding whitening matrix$\mathbf{W}'_s$. This provided what will be subsequently referred to as the classical calibration analysis:
  \begin{equation}\label{equ:classical}
    \mathbf{W}'_s (\boldsymbol{\kappa}_s - \mathbf{U}_f\boldsymbol{\kappa}_{calibrator}) = \mathbf{W}'_s \boldsymbol{\kappa}_{0,s} + \mathbf{W}'_s\boldsymbol{\varepsilon}_{s},
  \end{equation}
  Equations (\ref{equ:raw}), (\ref{equ:model-fitting_reduction}), and (\ref{equ:classical}) are used in parallel for comparison of the three approaches.
  
  \subsection{Targets}
  We observed targets of opportunity that were easily accessible over the engineering time of the instrument.
  3 Ser (HDS 2143, HIP 74649) is a K0III star with a main-sequence stellar companion expected at a separation of $\approx 280$ mas and a contrast in the infrared of $\Delta H \approx 5$ according to the orbit determined by \citet{Horch2015}. The observations took place during the engineering time of the SCExAO instrument, starting 2019-03-21T13:40:00. Here, we acquire $308 \mu s$ exposures at 3,193 Hz for 40 minutes during its transit. The total rotation during this period was 35 degrees.\par
  In order to compare the approach with a standard calibration procedure, a calibrator star of the same spectral type and the same apparent magnitude was selected in order to observe it with the same exposure times and similar AO performance. Proximity on sky is also necessary in order to replicate the airmass and instrumental flexures and for it to be observed just before or after the target. These constraints and the subjective decision they entail are a large part of the problems caused by classical calibration. In our case, we observed 31 Boo, a G7III star with the same frame rate and exposure times for a duration of two minutes between 13:09 and 13:40 UTC on the same night.\par
\begin{figure}
      \centering
      \includegraphics[width=0.45\textwidth]{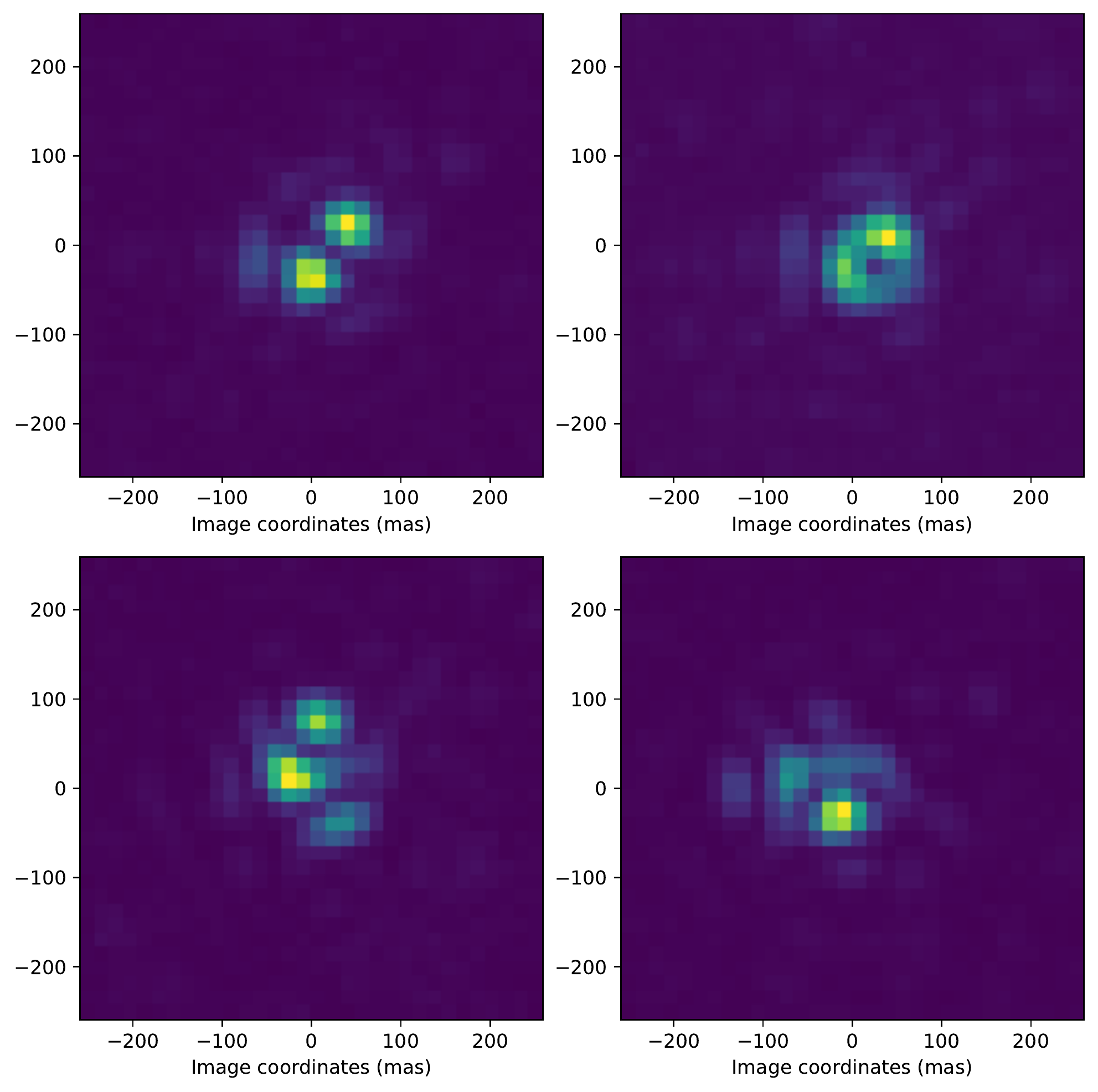}\par
      \vspace{5.pt}
      \includegraphics[width=0.45\textwidth]{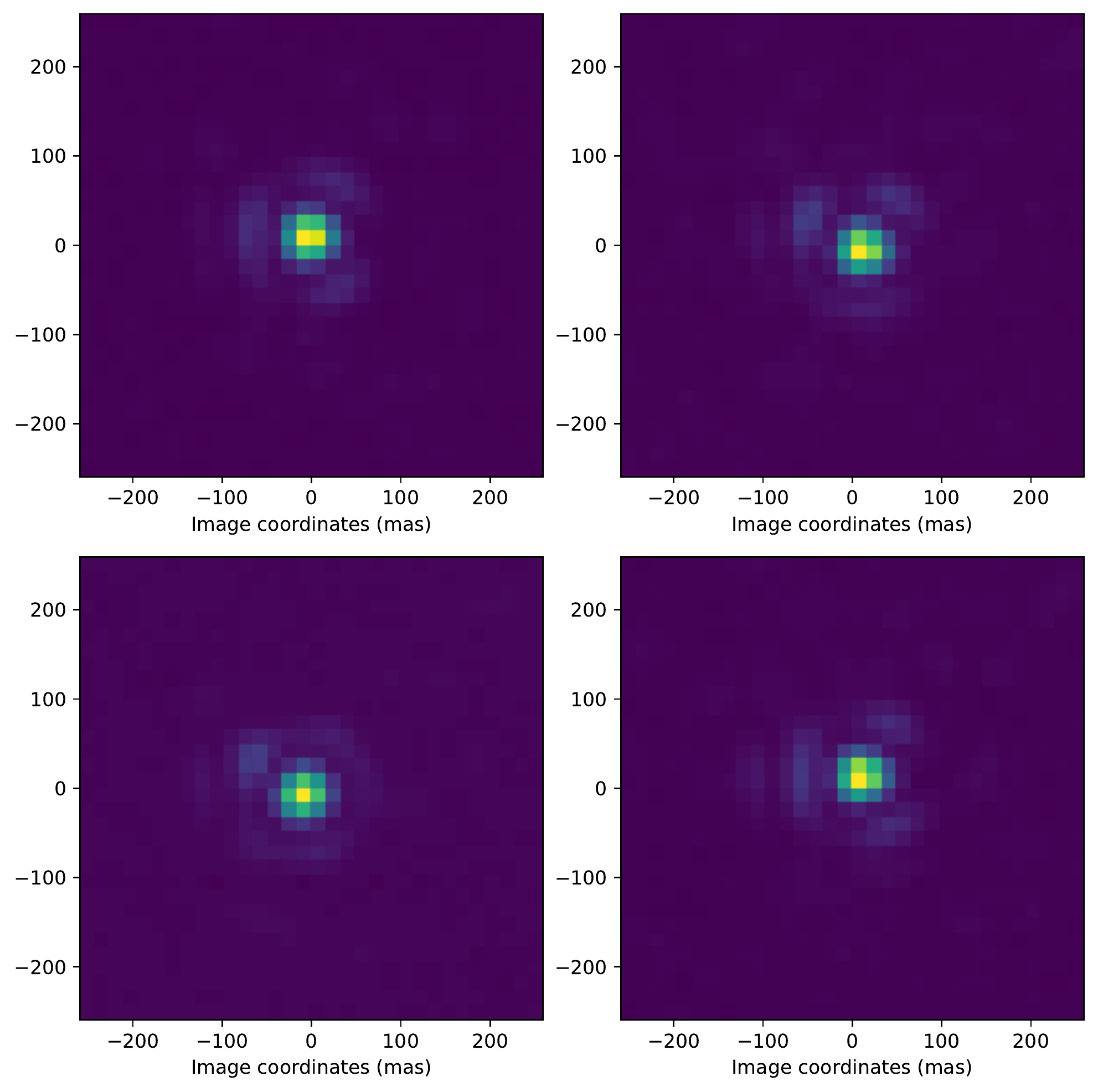}
      \caption{First two rows: images of the calibrator 31 Boo rejected at frame selection due to the low-wind effect. Last two rows: images kept for the analysis. Despite the selection, the appropriate images still display some variability on top of a static aberration revealed by the aspect of the first diffraction ring.}
      \label{fig:rejected-kept}
  \end{figure}

 
  During this observing session, the wind was low, resulting in a high prominence of low wind effect \citep{Sauvage2016a} with devastating consequences for the PSF at high spatial frequencies and evolutions in the timescales of seconds. This was dealt with during the processing through frame selection by selecting only the 5\% best frames. Figure \ref{fig:rejected-kept} shows a few examples of images of the calibrator 31 Boo which were kept and rejected at this stage. In our case, for a companion located at the expected separation of 280~mas, the kernel phases were binned into steps of 1.8 degrees. Some of those bins that contained too few images had to be discarded, leaving 11 bins covering a total field rotation of 29 degrees.\par
  For the second observing session, the aim was to study the stability of the instrument by observing a single star for a long period during its transit in a more typical set of observing conditions. We observed HD 211976 (HIP 110341) during an engineering time starting at 2019-06-25T14:21:00 for a duration of 38.5 minutes, resulting in a total field rotation of 40 degrees. Because the star was fainter, the integration time was pushed to 2ms at 500 Hz, and co-added four by four before a selection of the 50\% best images based on empirical Strehl ratio. The field rotation binning was kept at 1.8 degrees, providing 23 bins with a total field rotation of 40 degrees.\par
  Parameters for both targets are available in Table \ref{tab:extraction_parameters}.

  \subsection{Basic analysis}
\begin{figure}
        \centering
        \includegraphics[width=0.45\textwidth]{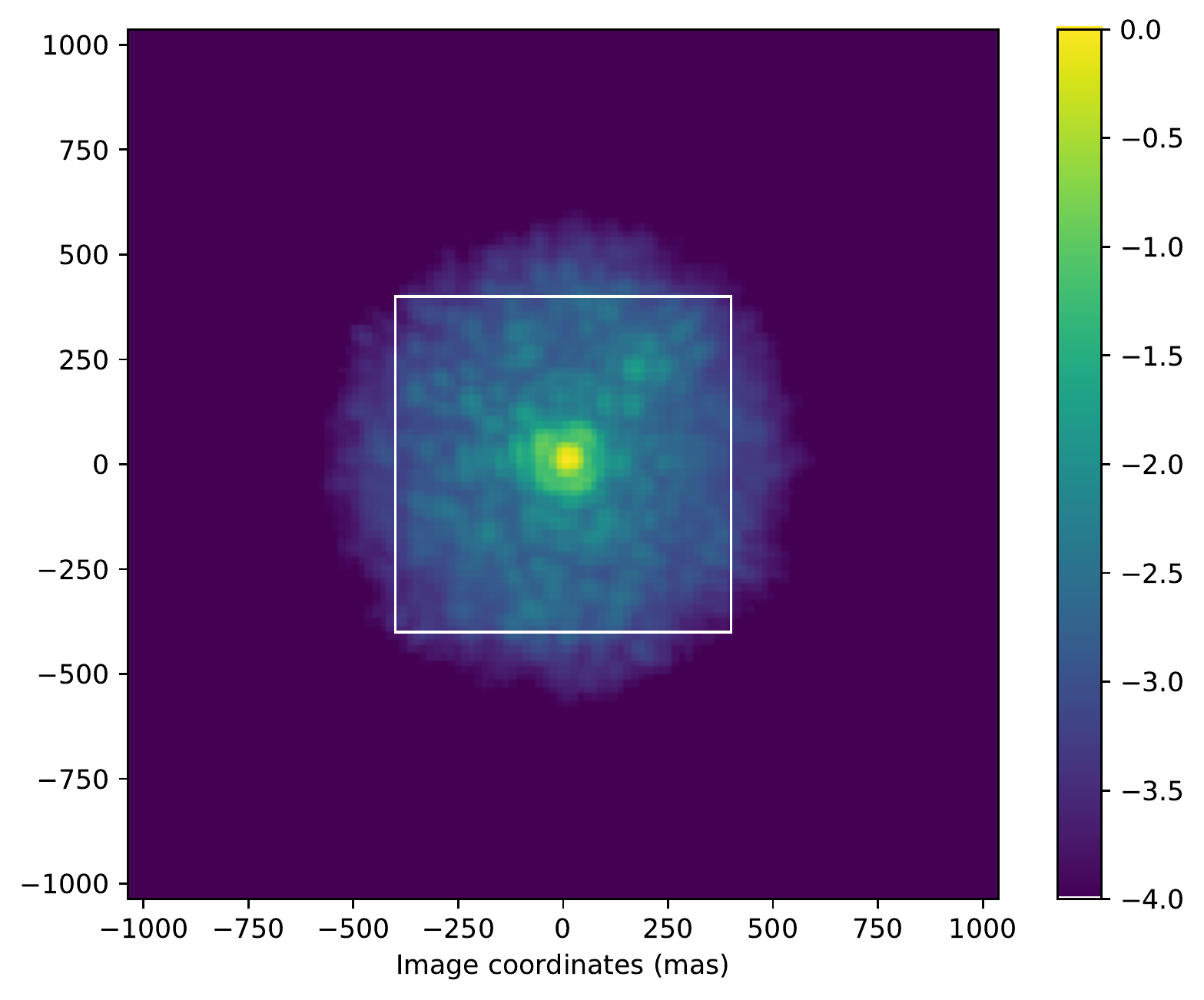}
        \caption{Lucky imaging integration extracted from the parallactic angle bin with PA $\approx -0.4\deg $. The white square represents the region of interest of the analysis and the axis are labelled in mas.}
        \label{fig:lucky_image}
    \end{figure}

  Lucky imaging using a stack of the images around parallactic angle -0.4 degrees is given in Fig. \ref{fig:lucky_image}, showing the instrumental PSF and the region of interest. Careful examination of the upper right quadrant reveals the companion hidden in a field of speckles.\par
  Colinearity maps are plotted as in \citet{Laugier2019} to highlight the possible location of the companion. The maximum value of this map is then used as a starting point for the extraction of the parameters of the binary by maximizing the likelihood with a Levenberg-Marquardt algorithm.\par
\begin{figure*}
        \centering
        \includegraphics[width=1.\textwidth]{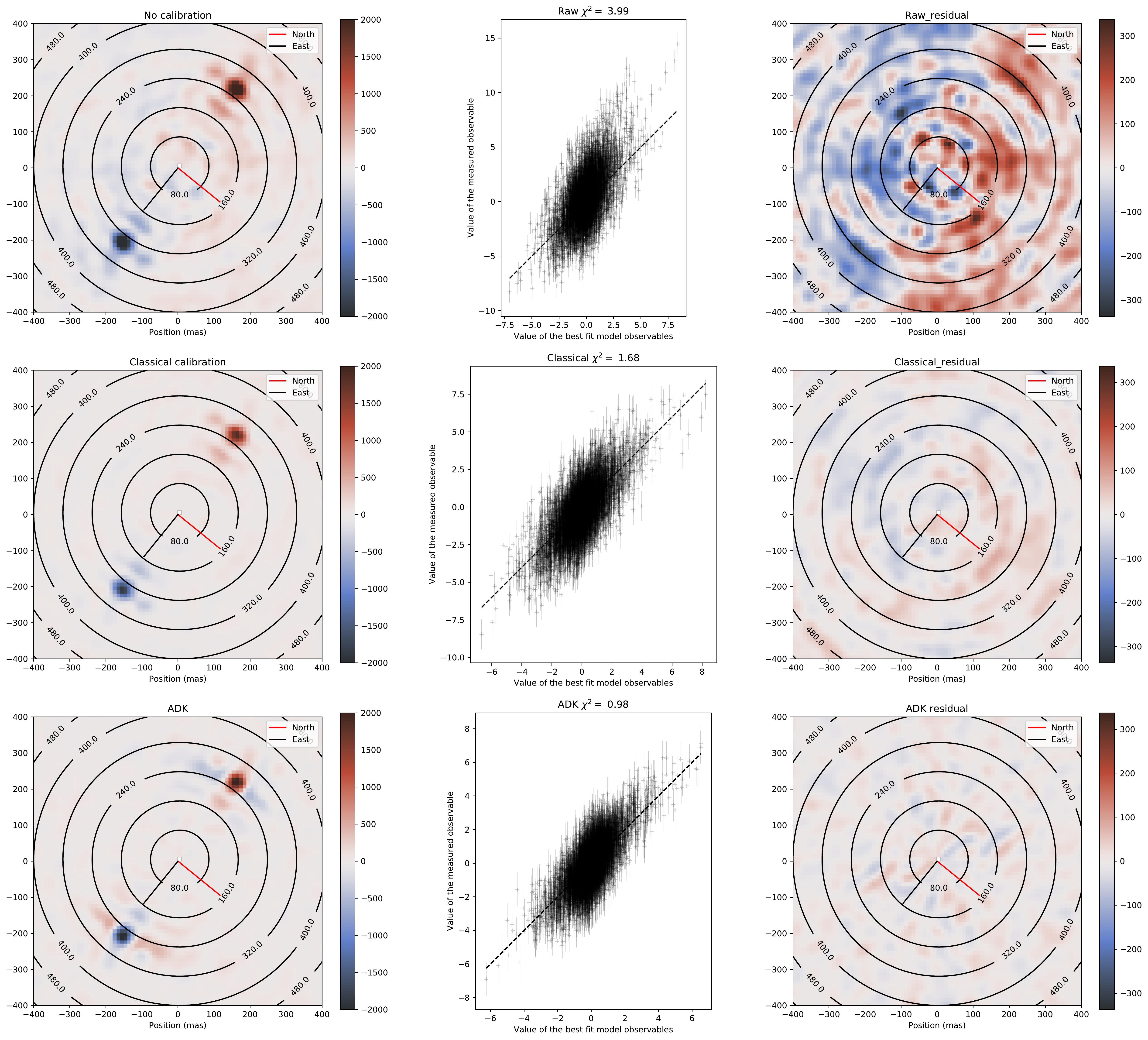}
        \caption{Left: colinearity maps of the region of interest around the target 3 Ser}. The orientation corresponds to the orientation on the detector at a parallactic angle of 0 deg. Middle:  correlograms corresponding to the final fit with adjusted covariance matrix. The corresponding $\chi^2$ is indicated for each plot. Right: colinearity map of the residual, after subtraction of the fitted binary signal. No additional companion is visible. The raw data has a significant amount of residual visible in the background and residual of the colinearity maps, as well as on the residual map. The level of bias is also visible with the poor fitting quality of the fit. ADK and classical calibration show comparable amount of residual background in the colinearity maps, but the $\chi^2$ value closer to 1 seem to indicate an even lower level of bias with ADK.
        \label{fig:grouped_maps}
    \end{figure*}

  Figure \ref{fig:grouped_maps} shows the colinearity maps where a companion is clearly visible, compatible with HDS 2143B. The companion's position and contrast are evaluated by model-fitting and colinearity maps of the residual of this fit shown in Fig. \ref{fig:grouped_maps} do not reveal any other strong binary signal.
  
\subsection{Re-evaluation of the covariance}
  The amplitude of the residuals of this model-fitting are much higher than expected. They do not show a strong binary signature, as revealed by colinearity maps of the residuals. We introduce the hypothesis that the extracted companion signal is the only significant kernel-phase signal on 3 Ser. Based on this assumption, we evaluate the residual variance of the observables along the course of the observation. This variance is then added to the diagonal of the covariance matrix of each slice in order to better represent the errors. In the case of the classical calibration, the signal of the calibrator was appended to the series in order to account for the evolution of the signal in the relevant timescale.\par
  Throughout the entire process, the different calibration strategies were applied independently from each other so that the residuals would not be affected by small differences in the model-fitting result.

\section{Results}
\subsection{Revised analysis}
Model fitting was performed anew with the corrected covariance estimation and the result is shown in Table \ref{tab:model_fit}. The results show consistent parameters for the three reduction pipelines. The estimated standard deviation for the position angle is significantly smaller in the case of ADK.\par
In the case of classical calibration, the higher value of $\chi^2$ is a sign of under estimation of the errors. This may indicate that a source of error was not taken into account. In this case, evolution of the bias between the calibrator and the target is a probable cause. This uncertainty is a typical source of problems in calibrating on-sky observations.
\begin{table}
    \centering
    \begin{tabular}{c|c|c|c}
    Mode & $\rho$ (mas) & $\theta$ (deg, E of N) & contrast \\
        \hline
        Raw & $265.1\pm0.3$ & $272.52\pm0.08$ & $58.\pm1.$ \\
        Classical & $266.8\pm0.2$ & $272.55\pm0.06$ & $65.1\pm0.8$\\
        ADK & $266.8\pm0.2$ & $272.6\pm0.04$ & $62.2\pm0.7$\\
    \end{tabular}
    \caption{Extracted binary parameters compared for the different calibration approaches.}
    \label{tab:model_fit}
\end{table}{}

\subsection{Performance comparison}
\begin{figure}
    \centering
    \includegraphics[width=0.45\textwidth]{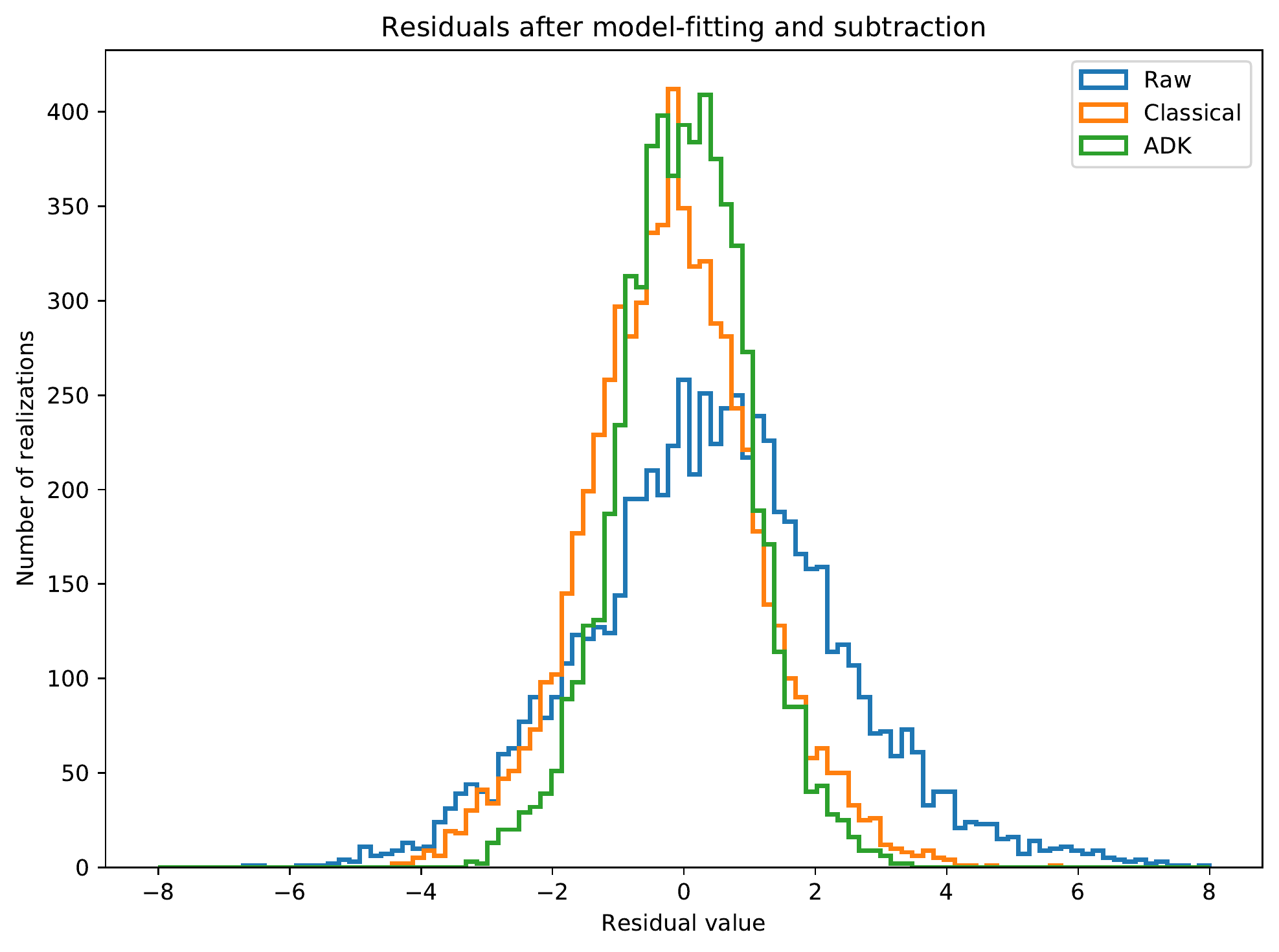}
    \caption{Comparison of the distribution of residuals after the final model fitting. The residual from classical calibration appears skewed to the left, whereas the ADK residual looks very compatible with a centered Gaussian distribution.}
    \label{fig:residual_distribution}
\end{figure}

The distribution of the residuals shown in Fig. \ref{fig:residual_distribution} demonstrates a comparable distribution between ADK and classical calibration residuals. \par

\begin{figure}
    \centering
    \includegraphics[width=0.45\textwidth]{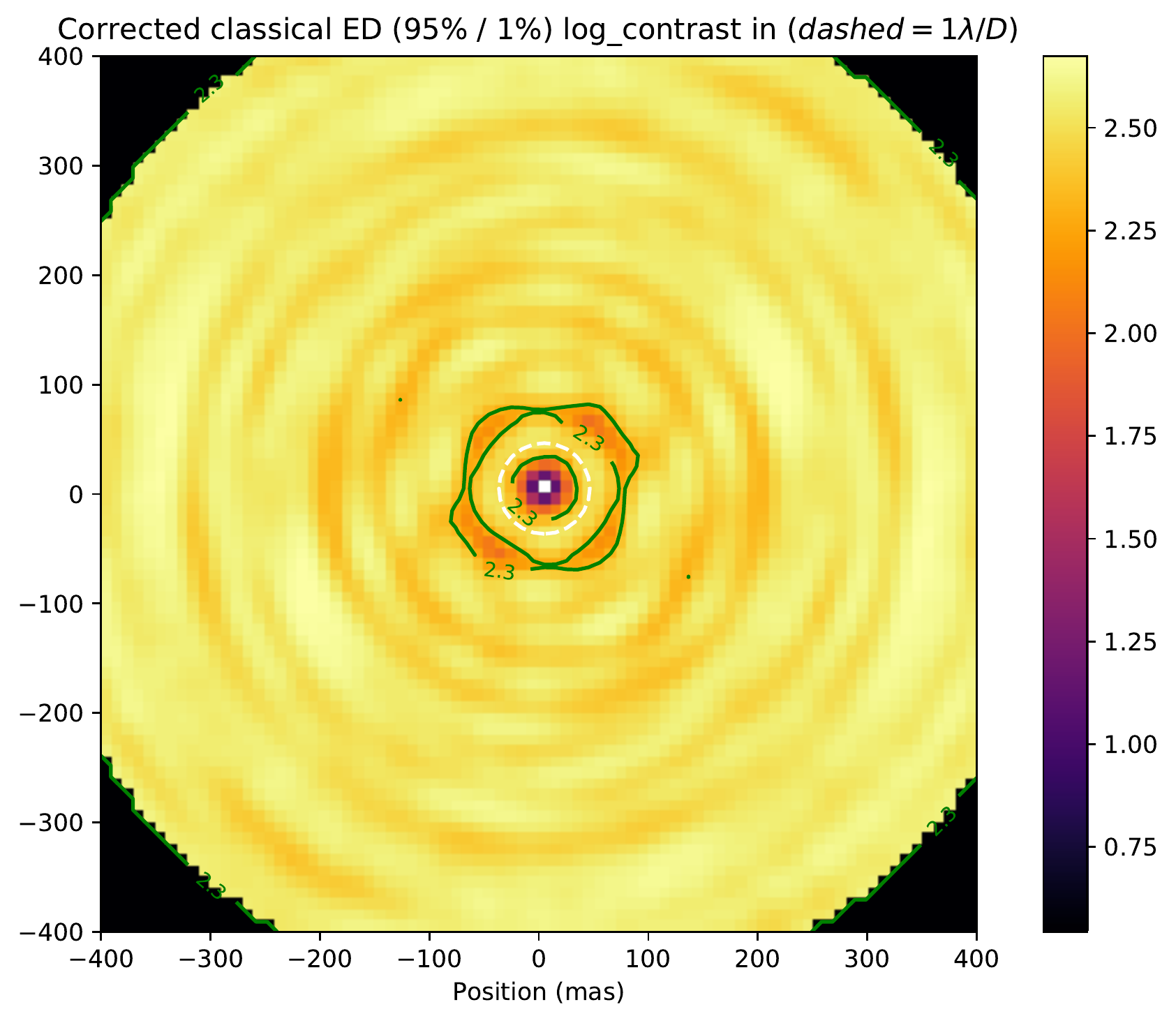}\par
    \includegraphics[width=0.45\textwidth]{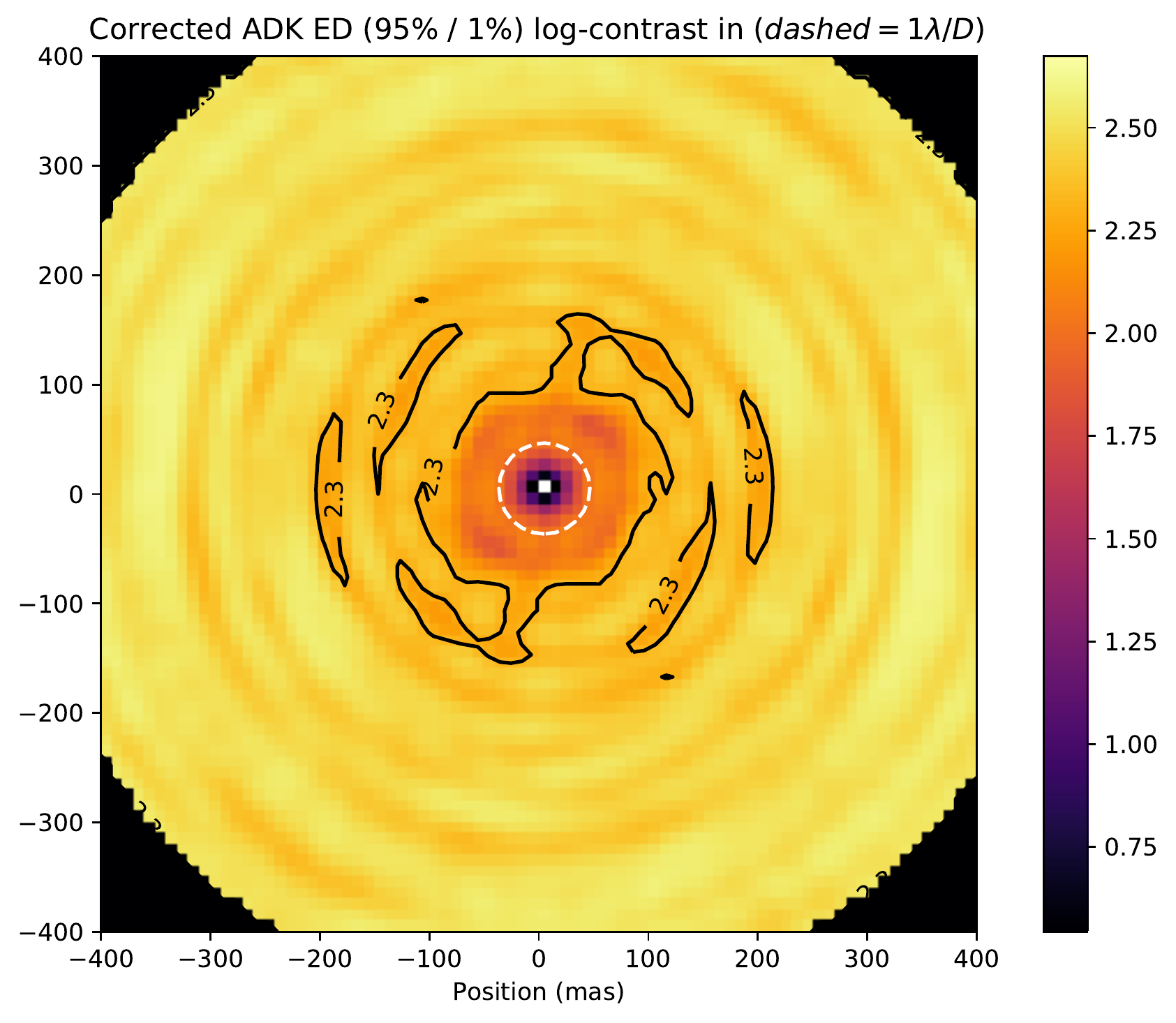}
    \caption{Classical calibration and ADK sensitivity maps compared. The sensitivity is expressed in log-contrast on a common scale and the dashed line marks a separation of 1$\lambda/D$. The performance in classical calibration are expected to be optimistic, as hinted by a larger value of $\chi^2 =1.68$.} 
    \label{fig:detection_maps}
\end{figure}

Based on the revised covariance estimation, sensitivity maps were drawn following the energy detector $T_E$ test proposed by \citet{Ceau2019b}. They were determined for a detection probability of 0.95 and false alarm rate of 0.01 for the region of interest, which is plotted in Fig. \ref{fig:detection_maps}. They show a sensitivity lower in ADK than in classical calibration, especially at small separations. Although sensitivity loss at small separation is expected in ADK, a significant part of the overall lower performance shown  here can be attributed to the under-evaluated covariance in the case of the classical calibration.\par

\subsection{Stability analysis}
\begin{figure}
    \centering
    \includegraphics[width=0.45\textwidth]{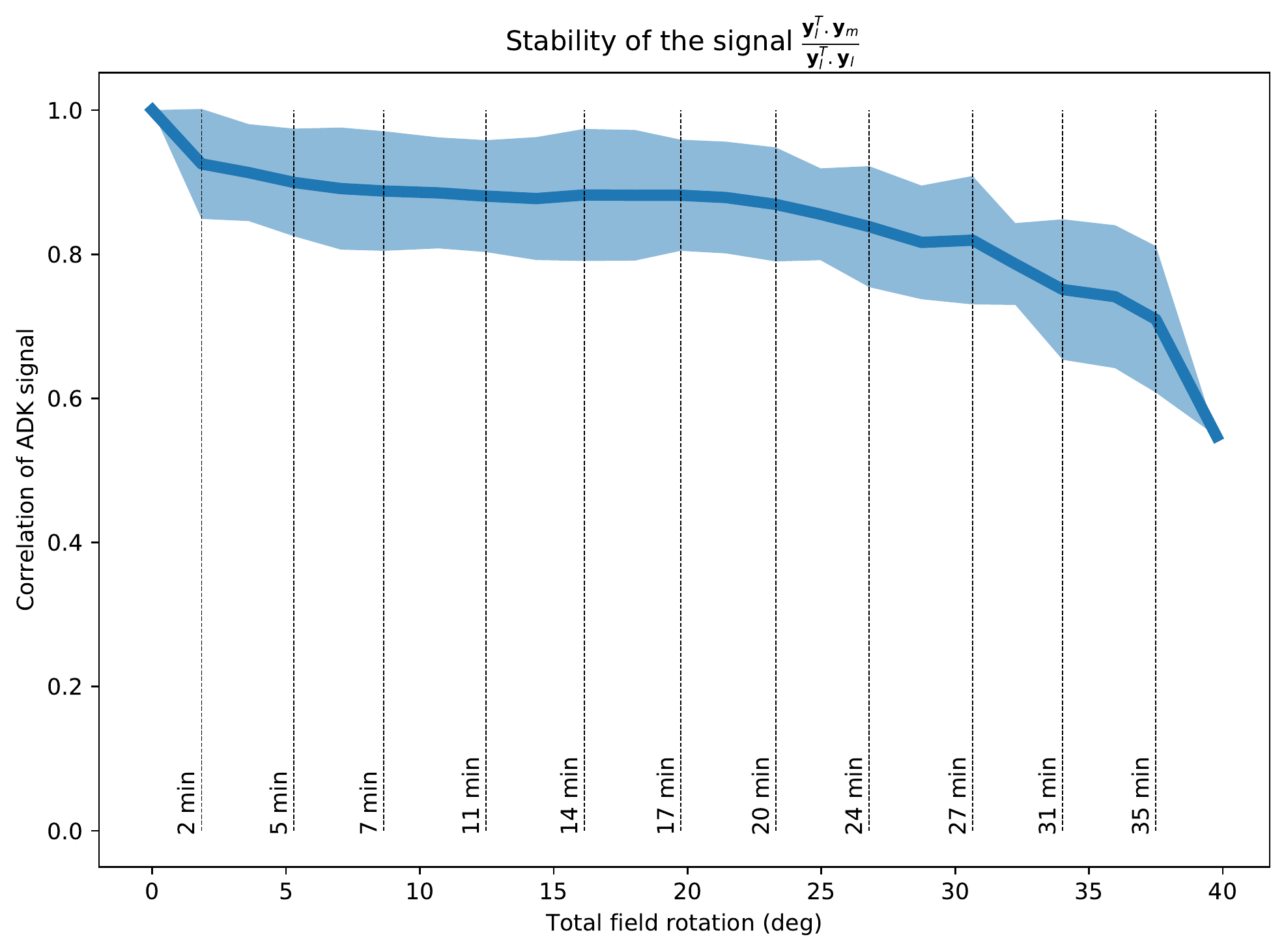}\par
    \includegraphics[width=0.45\textwidth]{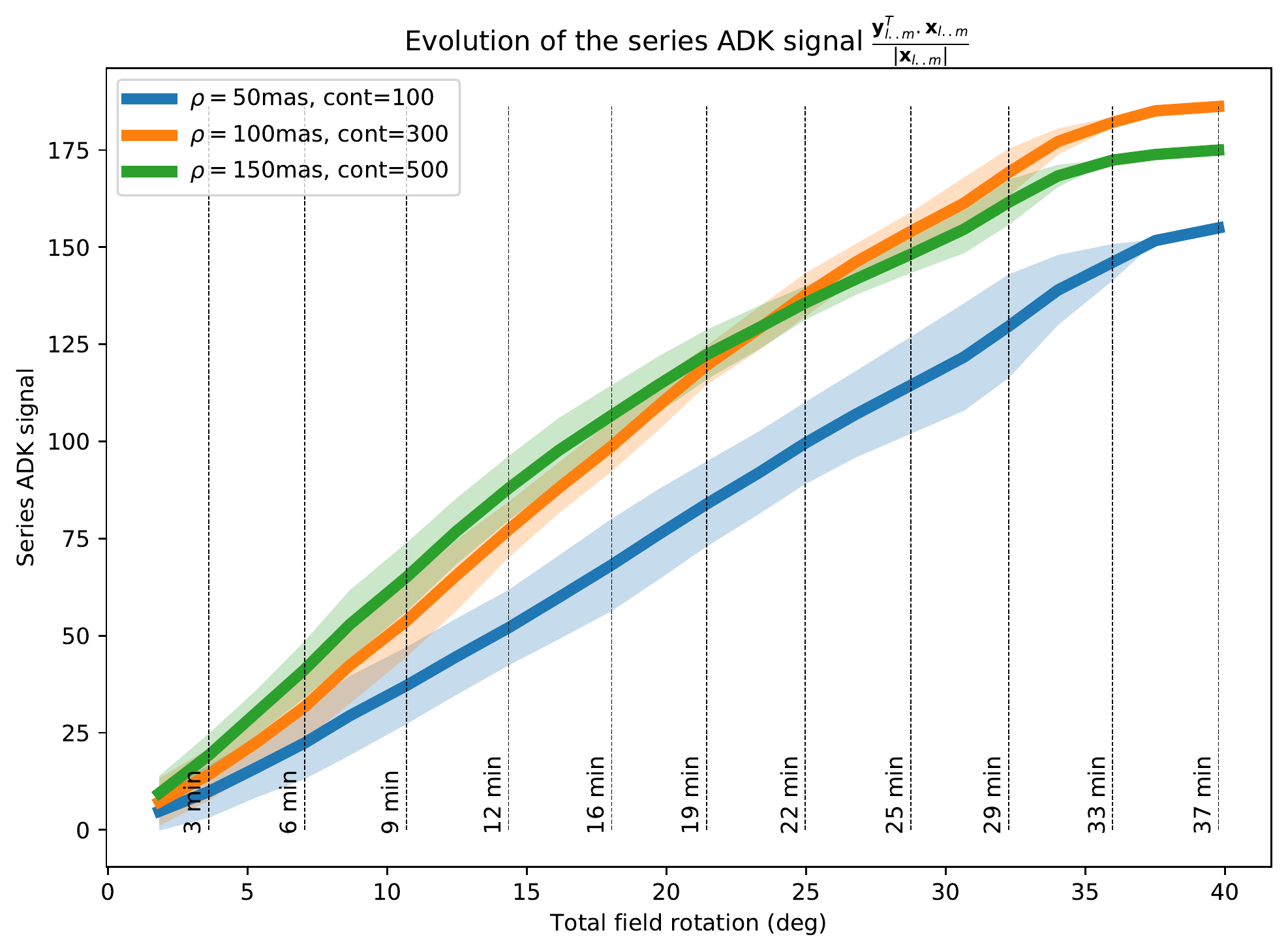}
    \caption{Representations of the evolution of the biases and signals with field rotation and time based on the observations of the single star HD 211976. Top: correlation between pairs of the kernel-phase vectors evolving with their temporal separation. Bottom: series ADK signal for different target separations. The colored regions represent the standard deviation of the realizations in our sample, therefore, it is not relevant for the largest field rotation samples that only have one realization. The contrast for the binary signals are adjusted to provide signals of similar amplitudes.}
    \label{fig:signal_evolution}
\end{figure}

Stability of the measurement is key to any calibration process. In classical calibration, the bias must remain static over the timescale necessary to switch from the calibration reference to the target, whereas in ADK, it must remain static during the necessary time to allow sufficient field rotation. We made use of the measurements on the single star HD 211976 to evaluate the evolution timescale of the bias signal. Since it is a single star, the only signal present is the bias signal. We considered each possible pair of kernel-phase signals $\mathbf{\kappa}_l$ and $\mathbf{\kappa}_m$, and built a common whitening matrix based on the mean of the two covariance matrices $\boldsymbol{\Sigma}_l$ and $\boldsymbol{\Sigma}_m$:
\begin{equation}
    \mathbf{W}_{l,m} = \Bigg( \frac{\boldsymbol{\Sigma}_l + \boldsymbol{\Sigma}_m}{2} \Bigg)^{-\frac{1}{2}}.\end{equation}
This defines the preferred basis in which we can compare the two vectors. Then we computed the correlation $c$ of the two vectors in this base:
\begin{equation}
    C = \frac{\mathbf{y}_l^T\cdot\mathbf{y}_m}{ \mathbf{y}_l^T\cdot\mathbf{y}_l},
\end{equation}{}
where $\mathbf{y}_l = \mathbf{W}_{l,m}\cdot\boldsymbol{\Sigma}_l$ and $\mathbf{y}_m = \mathbf{W}_{l,m}\cdot\boldsymbol{\Sigma}_m$. This correlation value is plotted in mean and standard deviation in Fig. \ref{fig:signal_evolution}. It shows that the signal loses the first 10\% of consistency on the 5 minutes timescale. However, 90\% of consistency is maintained over 20 minutes and around 80\% over 30 minutes. Although the sample is too small to draw conclusions at longer timescales, the stability seems to deteriorate after that point.\par
In the case of ADK, the bias must remain static in the timescale necessary for the field rotation to generate a significant signal in the differential kernel phases. As a consequence, the ADK sensitivity becomes the result of a race between the decorrelation of the signal and the decorrelation of the bias. This result is mostly influenced by the separation and contrast of the target and by the peak elevation of the target as seen from a given observatory (as it conditions the rate of field rotation). The amplitude of the bias signal is best evaluated in comparison to the signal of interest. Here, we add to the kernel-phase value of each bin, the corresponding theoretical signal $\boldsymbol{\kappa}_{s,th}$ of a companion. Then we produce a reduction of the signal similar to what is described in \ref{sect:reduction}, for all the possible pairs of bins -- as per the pairwise solution proposed with Equation (\ref{equ:pairwise_matrix}) -- and for all the possible series of consecutive bins (as per what is proposed in this work). We then project this signal onto the injected signal, in the same way as in the colinearity maps presented previously:
\begin{equation}
    M(l,m) = \frac{\mathbf{y}_{l..m}^T\cdot\mathbf{x}_{l..m}}{| \mathbf{x}_{l..m} |}
,\end{equation}{}
where 
\begin{equation}
    \mathbf{y}_{l..m} = \mathbf{W}_{L,l..m} \cdot \mathbf{L}'_{l..m}\cdot \Big(\boldsymbol{\kappa}_{s,l..m} + \boldsymbol{\kappa}_{s,th,l..m}\Big) 
\end{equation} and 
\begin{equation}
    \mathbf{x}_{l..m} = \mathbf{W}_{L,l..m} \cdot \mathbf{L}'_{l..m}\cdot \boldsymbol{\kappa}_{s,th,l..m}.
\end{equation}
The result, presented in Fig. \ref{fig:signal_evolution}, shows that the signal increases steadily during the acquisition and only starts to taper-off after 30 minutes when the bias decorrelates to a significant extent.\par
{}

\section{Discussion}
\subsection{Adverse conditions}
    In the comparative test, the performance of both methods are affected by two adverse effects that are contingent upon the observing conditions. The low wind effect introduces strong correlated wavefront errors and the use of frame selection introduces selection bias in the residual wavefront errors, which can dominate the quasi-static errors. These effects impact both the classical and ADK calibration procedures.\par
    In the case of both observing nights, the static non-common path aberrations are expected to degrade the rejection performance of the kernel-phase analysis as they impose a deviation from the linear regime. We expect this effect to increase the errors on the observables and, therefore, increase the covariance due to both quasi-static and quickly varying errors. This has an substantial effect on the sensitivity of the kernel-phase analysis in general.\par
    Because of the effect of these errors, our test does not provide a definitive reference for the performance of kernel-phase observations with SCExAO. However, the comparison between raw observables, classically calibrated observables, and ADK observables remains a topic of interest.\par
\subsection{Non-stationarity of the problem}
    In all of the test cases, despite the improvements brought by the ADK approach, our efforts at calibration were not sufficient to bring the residuals down to a level that could be explained by the covariances derived from each temporal bins. As in most of the published use-cases of robust observables, ad hoc adjustment was employed for the final analysis. This could be explained either by the presence of astrophysical signal different from a point-like companion (no response in the binary colinearity maps) or by the evolution of the bias at intermediate timescales.\par
    In the general case (not only in ADK), the rotation of the field of view introduces an added difficulty to the evaluation and treatment of this error because of the non-stationarity of the problem. The observables we use evolve in time, which imposes some constraints to the analysis. The temporal binning of the parallactic angle in particular must be selected with care as a compromise. Although its step has to be small enough that the signal of interest can be considered static in its timescale, a longer timescale may improve the quality of the evaluation of the covariance. The signature of binary stars decorrelates faster at larger separations, therefore requiring a finer temporal binning scheme than for smaller separations. Since the density of the pupil model also has to increase in order to reach larger separations, this may render explorations farther than $10 \lambda/D$ in long continuous sequences which are computationally demanding in ADK, where the projection matrix grows as $(n_k \cdot n_f)^2$.\par
    As is the case with ADI, the sensitivity is slightly reduced at very small separations, especially in the case where small field rotation is available.
    
\subsection{Conclusion}
    ADK is a new approach proposed for the calibration of robust observables that is aimed at removing most of the subjective choices that may often affect classical calibration techniques. It can be used both in the case of closure phases for non-redundant masking and in the case of      kernel phases for full-pupil imaging. In all our test cases, it was the approach that provided the smallest and most Gaussian residuals.\par
    Although our proposition for the matrix $\mathbf{L}'$ may seem mathematically simplistic compared to some of the more advanced forms of ADI, it was designed for the purposes of a flawless integration with our kernel-phase pipeline, especially with regard to the statistical whitening and hypothesis testing proposed by \citet{Ceau2019b}. Furthermore, the mathematical framework provided in appendix \ref{app:projection} can be used even with more elaborate definitions of the bias signal by replacing the matrix $\mathbf{U}_f$ with the appropriate matrices.\par
    The performance of the method currently appears to be limited by the opposition between the decorrelation of the signal of interest by field rotation and the decorrelation of the bias signal through the stability of the instrument. In the case of SCExAO, we showed that a steady accumulation of signal of interest was possible over more than 30 minutes and more than 33 degrees of field rotation, which makes the technique competitive even down to  $1\lambda/D$.\par
    Further improvements are expected with the coming improvements of the kernel-phase models, as well as the use of dedicated NCPA corrections for the internal infrared camera.

\begin{acknowledgements}
  KERNEL has received funding from the European Research Council (ERC) under the European Union's Horizon 2020 research and innovation program (grant agreement CoG - 683029).
  Based in part on data collected at Subaru Telescope, which is operated by the National Astronomical Observatory of Japan.
  The development of SCExAO was supported by the Japan Society for the Promotion of Science (Grant-in-Aid for Research \#23340051, \#26220704, \#23103002, \#19H00703 \& \#19H00695), the Astrobiology Center of the National Institutes of Natural Sciences, Japan, the Mt Cuba Foundation and the director's contingency fund at Subaru Telescope.
  The authors wish to recognise and acknowledge the very significant cultural role and reverence that the summit of Maunakea has always had within the indigenous Hawaiian community. We are most fortunate to have the opportunity to conduct observations from this mountain.
\end{acknowledgements}

   \bibliographystyle{aa} 
   \bibliography{biblio} 

\begin{appendix}

\section{The matrix L via projection}\label{app:projection}
 
There are several ways to obtain a matrix $\mathbf{L}$ that has the required properties. We will detail one and suggest a few approaches that eventually lead to mathematically equivalent results. We start with Equation (\ref{equ:uf}) describing the application of a static signal in the data:
\begin{equation}\label{equ:appendix_uf}
    \boldsymbol{\kappa}_s = \boldsymbol{\kappa}_{0,s} + \mathbf{U}_f \cdot \boldsymbol{\kappa}_{bias} + \boldsymbol{\varepsilon}_s,
\end{equation}
where $\mathbf{U}_f$ is the unfolding matrix described in \S \ref{sect:generalization}.\par
Considering that $\mathbf{U}_f$ is full column rank, the matrix $\mathbf{U}_f^\mathrm{T} \mathbf{U}_f$ is invertible, and a matrix $\mathbf{P}_U$ can be built that projects the signal into the space spanned by $\mathbf{U}_f$.
\begin{equation}
    \mathbf{P}_U =
    \mathbf{U}_f (\mathbf{U}_f^T \mathbf{U}_f)^{-1} \mathbf{U}_f^T.
\end{equation}
And conversely, a projection into the orthogonal complement to this subspace can be computed as:
\begin{equation}
    \mathbf{L} =
    \Big(\mathbf{I} - \mathbf{P}_{U}\Big).
\end{equation}
This matrix, therefore, projects the concatenated observables outside of the subspace reached by static bias as $ \mathbf{L} \mathbf{U}_f = 0$. This can easily be demonstrated as:
\begin{equation}
    \Big(\mathbf{I} - \mathbf{U}_f (\mathbf{U}_f^T \mathbf{U}_f)^{-1} \mathbf{U}_f^T\Big) \mathbf{U}_f =
    \mathbf{U}_f - \mathbf{U}_f (\mathbf{U}_f^T \mathbf{U}_f)^{-1} 
    \mathbf{U}_f^T\mathbf{U}_f,
\end{equation}
where $(\mathbf{U}_f^T \mathbf{U}_f)^{-1} \mathbf{U}_f^T\mathbf{U}_f = \mathbf{I}$.\par
In this particular case,
\begin{equation}
    (\mathbf{U}_f^T \mathbf{U}_f)^{-1} = \frac{1}{n_f}\mathbf{I},
\end{equation}
which leads us to write
\begin{equation}
    \mathbf{L} = \Big( \mathbf{I} - \frac{1}{n_f}\mathbf{U}_f  \mathbf{U}_f^\mathrm{T} \Big)
,\end{equation}
which is used in Equation (\ref{Lmatrix}).\par
It is fair to note a resemblance to an approach that would evaluate an estimate of $\boldsymbol{\kappa}_{bias}$, then subtract it from the observables of the series. Indeed, a rigorous least squares approach written in linear algebra, leads to the construction of the same matrix.\par
This reasoning would lead one to consider following this approach with a whitened observable $\mathbf{W}_s \cdot \boldsymbol{\kappa}_s$ in order to improve the estimate by benefiting from the weighting and decorrelation provided the whitening transform. This approach is provided in appendix \ref{app:post_whitening_proj} for reference, but it does not improve the result, since the subspace is exactly the same, as it is perfectly defined by the matrix $\mathbf{U}_f$.\par

\section{Managing the correlations}\label{app:dim_reduction}
Typical statistical analyses such as model fitting or detection tests rely on the computation of a likelihood functions which, for multivariate distributions, require the inversion of their covariance. In the typical kernel-phase analyses, this is commonly done by the construction of the whitening matrix. In both cases, this requires that the covariance matrix of the final observables be invertible. In the case of the basic $\mathbf{L}$ matrix, the covariance of the observables is written as:
\begin{equation}
    \textrm{Cov}(\mathbf{L}\boldsymbol{\kappa}_s) = 
    \mathbf{L}\textrm{Cov}(\boldsymbol{\varepsilon}_s)\mathbf{L}^T,
\end{equation}{}
where, in the best-case scenario, $\textrm{Cov}(\boldsymbol{\varepsilon}_s) = \mathbf{I}$, leads to:
\begin{equation}
    \textrm{Cov}(\mathbf{L}\boldsymbol{\kappa}_s) = 
    \mathbf{L}
\end{equation}
because $\mathbf{L}$ is symmetric ($\mathbf{L}^T = \mathbf{L}$) and, as a projection, idempotent ($\mathbf{L}^2 = \mathbf{L}$). Here $\mathbf{L}$, as the matrix of a projection, is not invertible. In the general case, this comes from the fact that $\mathbf{L}$ is non-surjective as it does not span the entire subspace of dimension $n_f \times n_k$. As a consequence, the observables cannot be directly used through the rest of the usual kernel-phase pipeline.\par
Since this is caused by the non-surjective nature of $\mathbf{L}$, an obvious solution to this problem is to reduce the dimensions of the output vectors to match the span of the matrix in order to make it into a surjection, and obtain an invertible covariance matrix. In the general case, an eigenvalue decomposition can be used to identify the subspace that is not reached by the transformation as will be defined by the eigenvectors that correspond to zero eigenvalues.\par
Here, we remark that the matrix is constituted of blocks corresponding to each frames, where each row of blocks can be expressed as a linear combination of all the other rows (minus their sum). We remove the last row and call this new matrix $\mathbf{L}'$. This reduces the number of rows to the rank of the matrix which is $(n_f-1)\times n_k  $, and therefore making it a surjection matrix.\par

\section{Computing the projection on whitened observables}\label{app:post_whitening_proj}

Since our construction of the matrix $\mathbf{L}$ is based around the expression of an estimator of $\boldsymbol{\kappa}_{bias}$, one might expect to obtain better result using a weighted estimation. In our case the error $\boldsymbol{\varepsilon}_s$ is correlated. A good way to deal with this is to use a whitening transformation $\mathbf{W}$ \citep{Ceau2019b} that provides both weighting and decorrelation in a single step. In the context of the concatenated observables, this translates to a block-diagonal $\mathbf{W}_s$ matrix that can be computed block by block as follows:
\begin{equation}
    \mathbf{W}_s = 
    \begin{bmatrix}
        \mathbf{W}_1 & 0 & \cdots & 0\\
        0 & \mathbf{W}_2 & \cdots & 0\\
        \vdots & \vdots & \ddots & \vdots\\
        0 & 0 & \cdots &  \mathbf{W}_{n_f}\\
    \end{bmatrix}
    = 
    \begin{bmatrix}
        \boldsymbol{\Sigma}_1^{-\frac{1}{2}} & 0 & \cdots & 0\\
        0 & \boldsymbol{\Sigma}_2^{-\frac{1}{2}} & \cdots & 0\\
        \vdots & \vdots & \ddots & \vdots\\
        0 & 0 & \cdots &  \boldsymbol{\Sigma}_{n_f}^{-\frac{1}{2}}\\
    \end{bmatrix}
\end{equation}
  where $\boldsymbol{\Sigma}_i$ is the covariance matrix of the $i^{th}$ kernel-phase observable vector. Working with this new observable, the steps are similar, although they are more complicated.
\begin{equation}\label{eq:revealing_kappacal}
\mathbf{W}_s \boldsymbol{\kappa}_s = \mathbf{W}_s \boldsymbol{\kappa}_{0,s} + \mathbf{W}_s \mathbf{U}_f \cdot \boldsymbol{\kappa}_{bias} + \mathbf{W}_s \boldsymbol{\varepsilon}_{sc}\,,
\end{equation}
where $\mathbf{U}_f$ is still the unfolding matrix that maps the constant calibrator signal into a series of repeated signals; and $ \boldsymbol{\kappa}_{bias} $ is the fixed calibrator signal.

As was done in appendix \ref{app:projection}, we can compute the matrix of the projection into the orthogonal complement of $\mathbf{U}_f$. This time, the method requires inverting $ \Big( \mathbf{W}_s \mathbf{U}_f\Big)^\mathrm{T}\mathbf{W}_s \mathbf{U}_f $ which is a bit more complicated, but can still be done very efficiently in the case where the errors between frames are independent ($\mathbf{W}_s$ is block-diagonal) because of the structure of both matrices. The result is a $ n_k$ by $ n_k $ matrix $\mathbf{B}$.
\begin{equation}\label{eq:l2.1}
    \mathbf{B}  = 
    \Bigg( \Big( \mathbf{W}_s \mathbf{U}_f\Big)^\mathrm{T}\mathbf{W}_s \mathbf{U}_f \Bigg)^{-1}
    = \Big( \displaystyle\sum_{i=1}^{n_f} \mathbf{W}_i^2 \Big)^{-1}
.\end{equation}
By defining the matrix as
\begin{equation}\label{eq:l2.2}
    \mathbf{L}_2 = \Big( \mathbf{I} - \mathbf{W}_s \mathbf{U}_f \mathbf{B} \mathbf{U}_f^\mathrm{T} \mathbf{W}_s \Big) ,
\end{equation}
we can simplify the equation into the following:
\begin{equation}\label{eq:l2.3}
   \mathbf{L}_2 \mathbf{W}_s \boldsymbol{\kappa}_s =
    \mathbf{L}_2 \mathbf{W}_s \boldsymbol{\kappa}_{0,s} + \mathbf{L}_2 \mathbf{W}_s \boldsymbol{\varepsilon}_s .
\end{equation}
This approach uses the hypothesis on the calibration signal to project the observables in a subspace that is beyond its reach. Now since the error term $ \mathbf{L}_2 \mathbf{W}_s \boldsymbol{\varepsilon}_s$ is correlated, there is an additional step to perform as its covariance $\mathbf{L}_2$ is not invertible. However, since $\mathbf{L}_2$ is a projection, we can build a different projection matrix $ \mathbf{L}_2' $ that has full column-rank. As a real symmetric matrix, $\mathbf{L}_2$ is diagonalized as per the form:

\begin{equation}
    \mathbf{L}_2 =  \mathbf{V} \boldsymbol{\Lambda} \mathbf{V}^{-1}  =  \mathbf{V} \boldsymbol{\Lambda} \mathbf{V}^\mathrm{T} ,
\end{equation}
where $\boldsymbol{\Lambda}$ is a diagonal matrix containing 0s and 1s (because $\mathbf{L}_2$ is a projection). In the subspace defined by $\mathbf{V}^\mathrm{T}$, the operation corresponds to cropping the observable vector by $n_k$. We can therefore define a new projection matrix with full column rank with the rows of $\mathbf{V}^\mathrm{T}$ corresponding to the non-zero eigenvalues of $\mathbf{L}_2$ (here by using $\boldsymbol{\Lambda}'$ which is a correspondingly cropped version of $\boldsymbol{\Lambda}$):
\begin{equation}\label{eq:l2.crop}
    \mathbf{L}_2' = \boldsymbol{\Lambda}' \mathbf{V}^\mathrm{T} .
\end{equation}
This time, the covariance of the error parameter $\mathbf{L}_2' \mathbf{W}_s \boldsymbol{\varepsilon}_s$ is:
\begin{equation}
    \boldsymbol{\Lambda}' \mathbf{V}^\mathrm{T} \mathbf{I} \Big( \boldsymbol{\Lambda}' \mathbf{V}^\mathrm{T} \Big)^\mathrm{T} = \mathbf{I} .
\end{equation}
 With this new projection matrix, Equation (\ref{eq:l2.3}) becomes:
\begin{equation}\label{eq:final_equation}
    \mathbf{L}_2' \mathbf{W}_s \boldsymbol{\kappa}_s =
    \mathbf{L}_2' \mathbf{W}_s \boldsymbol{\kappa}_{0,s} + \mathbf{L}_2' \mathbf{W}_s \boldsymbol{\varepsilon}_s .
\end{equation}

Once again, we have successfully built a new projected observable in a smaller subspace that is robust to static errors ($\mathbf{L}_2' \mathbf{W}_s \mathbf{U}_f \boldsymbol{\kappa}_{bias} = 0$) and has identity covariance ($\mathrm{Cov} \Big( \mathbf{L}_2' \mathbf{W}_s \boldsymbol{\varepsilon}_s \Big) = \mathbf{I}$).
This allows for the construction of a likelihood function for $\boldsymbol{\kappa}_{0,s}$ and the application of all the statistical tools defined in \cite{Ceau2019b}. In other terms, it means that based simply on the covariance $\boldsymbol{\Sigma}_i$ of the observables in the different frames, we can easily build the matrix $ \mathbf{L}_2'$ that transforms Equation (\ref{eq:revealing_kappacal}) into Equation (\ref{eq:final_equation}), effectively removing the bias signal; and because of the precautions taken here, it can be used in any model fitting or statistical test as a drop-in replacement of usual whitened observables.\par
The matrices $\mathbf{L}'$ and $\mathbf{L}_2'$ are both designed to explore the same subspace that is purely defined as the orthogonal complement of the subspace of $\mathbf{U}_f$. Although the observables are described in a different basis, the likelihood function that they provide is expected to be the same.\par
A transfer matrix $\mathbf{M}$ can be computed to pass from one set of observables to the other:
\begin{equation}
    \mathbf{M} = \mathbf{L}_2'\mathbf{W}_s \Big( \mathbf{W}_L \mathbf{L}' \Big)^+ ,
\end{equation}{}
where the + sign indicates a Moore-Penrose pseudo-inverse. Although we do not provide a proof for it, we have found in our cases that $\mathbf{M}$ is a real unitary matrix which both confirms that they span the same subspace and that they provide the same sensitivity since they produce the same $\chi^2$:
\begin{equation}
    \chi^2\Big( \mathbf{L}_2'\mathbf{W}_s \boldsymbol{\kappa}_s \Big) = 
                    \chi^2\Big( \mathbf{M}\mathbf{W}_L \mathbf{L}' \boldsymbol{\kappa}_s \Big),
\end{equation}
which develops:
\begin{equation}
    \chi^2\Big( \mathbf{L}_2'\mathbf{W}_s \boldsymbol{\kappa}_s \Big) =
            \Big( \mathbf{M} \mathbf{W}_L \mathbf{L}' \boldsymbol{\kappa}_s \Big)^T \mathbf{M} \mathbf{W}_L \mathbf{L}' \boldsymbol{\kappa}_s
\end{equation}
and since $\mathbf{M}^T \mathbf{M} = \mathbf{I}$, we obtain:
\begin{equation}
    \chi^2\Big( \mathbf{L}_2'\mathbf{W}_s \boldsymbol{\kappa}_s \Big) = 
                    \chi^2\Big( \mathbf{W}_L \mathbf{L}' \boldsymbol{\kappa}_s \Big).
\end{equation}

\end{appendix}

\end{document}